\documentclass[useAMS,usenatbib,draftcopy]{mn2e}

\usepackage{graphicx}
\usepackage[light]{}

\def\kms      {\ifmmode{\rm km\,s}^{-1} \else km\,s$^{-1}$\fi}
\def\mujybm{\ifmmode{\rm \mu Jy}\,{\rm beam}^{-1}\else${\rm \mu}$Jy\,beam$^{-1}$\fi}
\def\ltsim{\ifmmode\stackrel{<}{_{\sim}}\else$\stackrel{<}{_{\sim}}$\fi}
\def\gtsim{\ifmmode\stackrel{>}{_{\sim}}\else$\stackrel{>}{_{\sim}}$\fi}
\def\S4195{41.95+575}
\def\S4331{43.31+592}
\begin{document}  
\title[SNR in M82]{15 years of VLBI observations of two compact radio sources in Messier 82}

\author[Beswick {\it et al.}]{R.\,J.\,Beswick,$\!\!^1$\thanks{Robert.Beswick@manchester.ac.uk} J.\,D.\,Riley,$\!^{1,2}$ I.\,Marti-Vidal,$\!^{1,3}$
A.\,Pedlar,$\!^1$ T.\,W.\,B.\,Muxlow,$\!^1$  
\newauthor  
A.\,R.\,McDonald,$\!^1$ K.\,A.\,Wills,$\!^4$ D.\,Fenech,$\!^1$ \& M.\,K.\,Argo$^{1}$\\
 $^1$Jodrell Bank Observatory, The University of Manchester, Macclesfield, Cheshire, SK11~9DL\\
 $^2$Centre for Astrophysics, University of Central Lancashire, Preston, Lancashire, PR1~2HE\\
 $^3$Departament d' Astronomia, Universitat de Val{\`e}ncia, 46100 Burjassot, Spain\\
$^4$Department of Physics and Astronomy, University of Sheffield, Hounsfield Road, Sheffield, S3~7RH}
 
\date{Accepted 2006 March 22.  Received 2006 March 21; in original form 2006 
February 16}
\pagerange{\pageref{firstpage}--\pageref{lastpage}} \pubyear{2006}

\maketitle
\label{firstpage}

\begin{abstract} {
 
We present the results of a second epoch of 18\,cm global Very Long-Baseline 
Interferometry (VLBI) observations, taken on 23 February 2001, of the central 
kiloparsec of the nearby starburst galaxy Messier 82. These observations further 
investigate the structural and flux evolution of the most compact radio sources 
in the central region of M82. The two most compact radio objects in M82 have been 
investigated (41.95+575 and 43.31+592). Using this recent epoch of data in 
comparison with our previous global VLBI observations and two earlier epochs of 
European VLBI Network observations we measure expansion velocities in the range 
of 1500--2000\,km\,s$^{-1}$ for 41.95+575, and 9000--11000\,km\,s$^{-1}$ for 
43.31+592 using various independent methods. In each case the measured remnant 
expansion velocities are significantly larger than the canonical expansion 
velocity (500\,km\,s$^{-1}$) of supernova remnants within M82 predicted from 
theoretical models.

In this paper we discuss the implications of these measured expansion
velocities with respect to the high density environment that the SNR are expected
to reside in within the centre of the M82 starburst.}

\end{abstract}

\section{Introduction}

Studies of extra-galactic supernova remnants (SNR) are currently
limited by sensitivity, and to some extent angular resolution,
nevertheless such studies provide unique insights into the early
evolution of radio supernovae and supernova remnants. The
extensive studies of remnants in our own galaxy ({\it e.g.} \citealt{green04}) are
particularly valuable for testing details of the physical
processes occurring in individual remnants ({\it e. g.}
Braun, Gull \& Perley 1987; \citealt{brogan06}\nocite{braun87}), but they are limited
by the fact that the youngest known galactic supernova remnant is
over 300 years old and that the distances to galactic remnants, in many
cases, remains quite unknown. 

Studies of extra-galactic SNR in starburst galaxies have a number
of advantages. Firstly, the high star-formation rate results in a
relatively large number of young supernova remnants with ages
measured in decades, rather than centuries. Secondly, as the
starburst region is typically a kiloparsec in extent, and as
distances to even the nearest starburst galaxies are a few Mpc,
the relative distances to each SNR will only vary by $\sim$0.1\,\%.
Finally, radio synthesis techniques ensure that all the SNR, within an
individual starburst galaxy,
are observed with same angular resolution and sensitivity.  Hence
as all the SNR are essentially at the same distance this
corresponds to a sample of SNR measured with constant linear
resolution and surface luminosity limit.

\begin{table*}
\centering
\caption{Summary of VLBI observations of M82.}
\label{tab1}       
\begin{tabular}{ccccl}
\hline\noalign{\smallskip}
Epoch & Observation date & Array & Frequency & Reference \\
 &  &   &  & \\
                                                                                
\hline
1&	11 Dec 1986 (1986.95)&Ef, Jb, Wb, Mc&1.4\,GHz&Pedlar {\it et al.,} 1999\\
2&	02 Jun 1997 (1997.42)&Ef, Jb, Mc, Nt, On, Wb, Cm, Tr&1.6\,GHz&Pedlar {\it et al.,} 1999\\
3&	28 Nov 1998 (1998.91)&VLBA, Y1, Ro, Go, Ef, Jb, Mc, Nt, On, Wb, Tr&1.6\,GHz&McDonald {\it et al.,} 2001\\
4&	23 Feb 2001 (2001.15)&VLBA, Y1, Ro, Ef, Jb, Mc, Nt, On, Wb, Tr&1.6\,GHz&This paper\\
\noalign{\smallskip}\hline
\end{tabular}
\end{table*}

\begin{figure*}
\begin{center}
\setlength{\unitlength}{1mm}
    \begin{picture}(80,75)
\put(0,0){\includegraphics{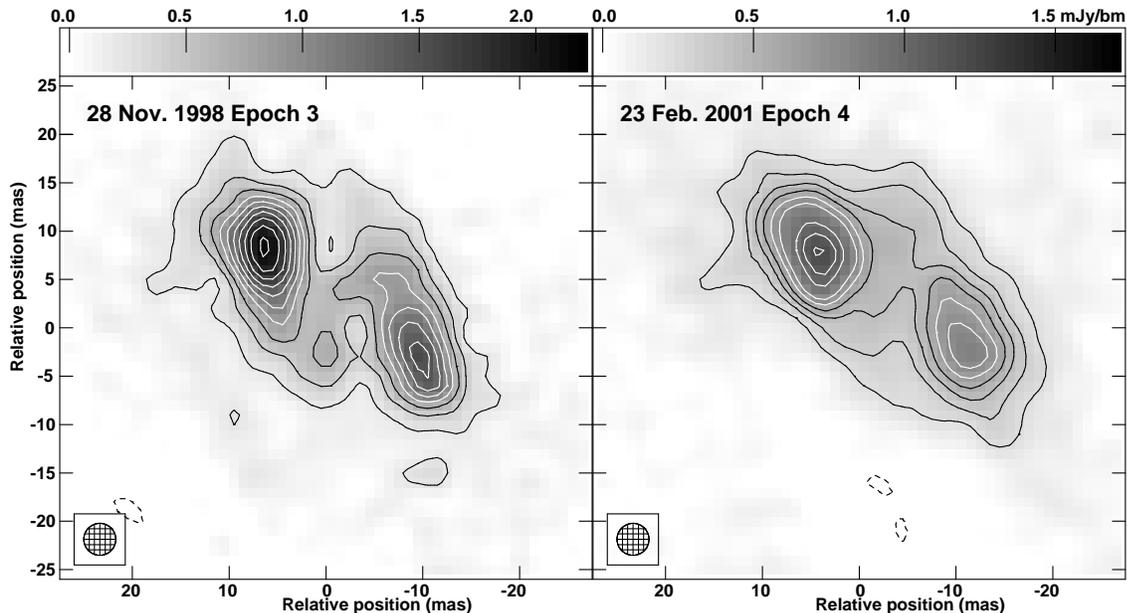}}
\end{picture}
\caption{Contour images of the compact radio source 41.95$+$575 from
the global VLBI epochs observed on 28 Nov. 1998 and 23 Feb. 2001. Both
images have been convolved with a 3.3\,mas circular beam. The two
epochs are contoured with linear multiples ($-$1, 1, 2, ... 10) times
0.21\,mJy\,beam$^{-1}$
 and 
 0.11\,mJy\,beam$^{-1}$ for the epochs 3 and 4 respectively.}
\label{fig1}
\end{center}
\end{figure*}

The star-formation rate (SFR) of a galaxy can be estimated using
measurements in the ultra-violet, optical, infra-red and radio regimes
({\it e.g.} \citealt{cram98}). However, as starburst galaxies usually contain 
large quantities of
dust, the estimates of SFR from ultraviolet and optical data are
uncertain due to dust extinction. M82 is no exception with
A$_{\rm v}$ typically 20--30 magnitudes \citep{mattila01}. Fortunately, radio
methods are not affected by dust extinction and hence the
SFR can be estimated from the total 1.4\,GHz
luminosity of the galaxy \citep{cram98}.  The total flux density of
M82 at 1.4\,GHz is $\sim$8.5\,Jy and using a distance of 3.2\,Mpc,
then the
SFR(M$\ge$5\,M${_\odot}$) for M82 can be estimated using
the relation given by Cram {\it et al.,} to be $\sim$2.5\,M$_{\odot}$\,yr$^{-1}$.  Also the total FIR
luminosity implies a value of SFR(M$\ge$5\,M$_{\odot}$)
close to 2\,M$_\odot$\,yr$^{-1}$ (\citealt{cram98,pedlar01}). In addition the
radio thermal free-free contribution is reasonably well defined in M82
and can be used to estimate a SFR (\citealt{pedlar99})
consistent with the above value. 

If the SFR is constant over $\sim10^7$ years, then in principle
the type-II supernova rate can be derived directly from the
SFR and the initial mass function (IMF), on the assumption
that all stars with masses $>$8M$_\odot$ become supernovae. Hence
assuming a Miller-Scalo IMF \citep{miller79} the supernova rate, $\nu_{sn}$, is
directly related to the SFR. This
implies a  $\nu_{sn}\sim$0.08$-$0.1\,yr$^{-1}$
in M82 and hence it is not surprising to find a relatively large
number of radio supernova remnants in this source \citep{muxlow94}.
\begin{figure*}
\begin{center}
\setlength{\unitlength}{1mm}
    \begin{picture}(80,82)
\put(0,0){\includegraphics{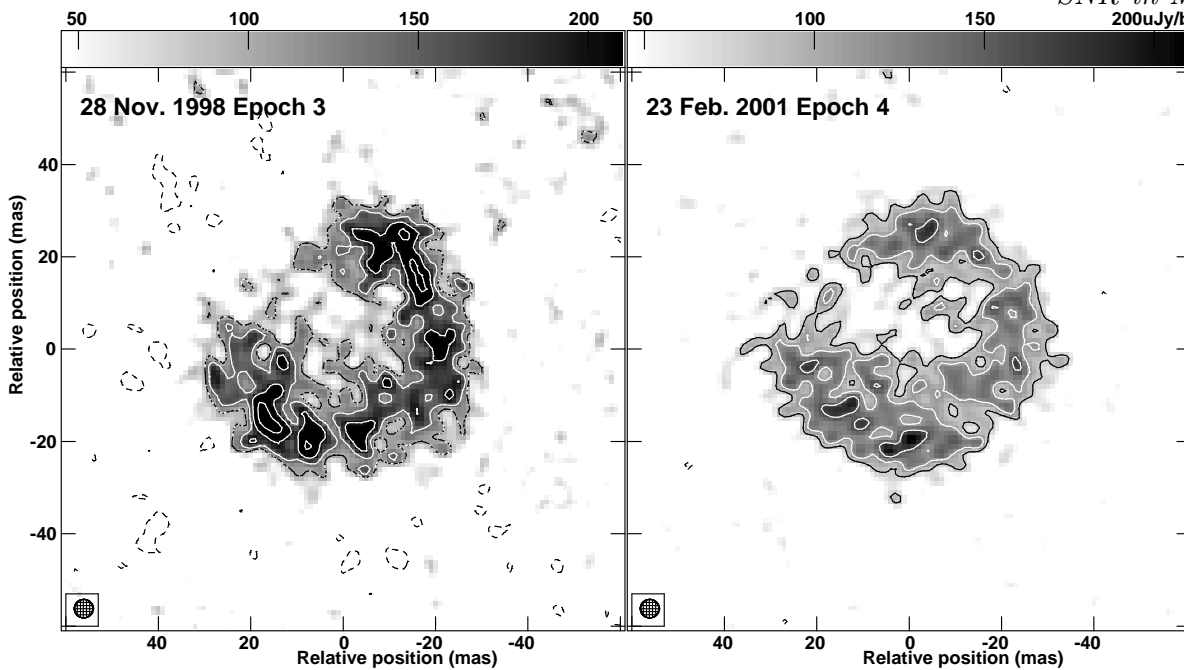}}
\end{picture}
\caption{Contour images of the  radio source 43.31$+$592 
from
the global VLBI epochs observed on 28 Nov. 1998 and 23 Feb. 2001. Both
images have been convolved with the same  4\,mas circular beam. The
two images have been contoured with multiples of $-$1, 1, 1.414, 2,
2.828, 4 \& 5.657 $\times$ 0.1\,mJy\,beam$^{-1}$ and
0.08\,mJy\,beam$^{-1}$ respectively.}

\label{fig2}
\end{center}
\end{figure*}

Several methods have been used to estimate the  $\nu_{sn}$
directly from the radio parameters of the M82 supernova remnants
(\citealt{muxlow94}; \citealt{pedlar99}). The simplest method is to assume
that all the M82 remnants which are brighter and smaller than the
Cassiopeia A are younger than 330 years.
Alternatively if a constant expansion velocity is assumed ({\it e.\,g.}~5000--10000\,km\,s$^{-1}$) this gives the ages of the
remnants and hence the $\nu_{sn}$ \citep{muxlow94}. Finally \citet{vanburen94} estimated the  $\nu_{sn}$
directly from remnant luminosities. Each of these methods imply
$\nu_{sn}$ between 0.05 to 0.1\,yr$^{-1}$, which is
consistent with the SFRs discussed above. 

These supernova rates imply that the $\sim$50 observed remnants
have ages typically of a few hundred years. However some doubts
have been raised concerning the evolution of the remnants in M82.
 Kronberg, Biermann \& Schwab (1985)\nocite{kronberg85} claim, using statistical arguments, that in a large
fraction of the remnants the radio luminosity is decaying at less
than $0.1 \, \%$ per year. If true this could suggest that the
remnants are over a thousand years old. Also theoretical
studies by \citet{chevalier01} have suggested
that the high pressure of the interstellar medium in M82 would
result in the supernova remnants expanding at only 500\,km\,s$^{-1}$, hence as the observed sizes are typically a
few parsecs, this again would imply ages of thousands of years.

In order to investigate these problems it is essential that the
expansion velocities and possible decelerations of the M82 
remnants. This is possible using Very Long Baseline Interferometry
(VLBI) techniques which enables their sub-parsec structure to be
measured. Pedlar {\it et al.} (1999) used the European VLBI network
(EVN) to produce the first VLBI images of the M82 SNR
with 20\,mas (0.3\,pc) resolution. This
work was followed up by \citet{mcdonald01} at even higher
angular resolution (5\,mas $\approx$0.1\,pc)
using the Global VLBI network.

In this paper we report on further global VLBI measurements taken over
two years after the McDonald~{\it et~al.} measurements.  They provide a
4th epoch of VLBI measurements and a 2nd epoch of global VLBI images of
these SNR.

\begin{figure*}
\begin{center}
\setlength{\unitlength}{1mm}
    \begin{picture}(80,48)
\put(0,0){\includegraphics{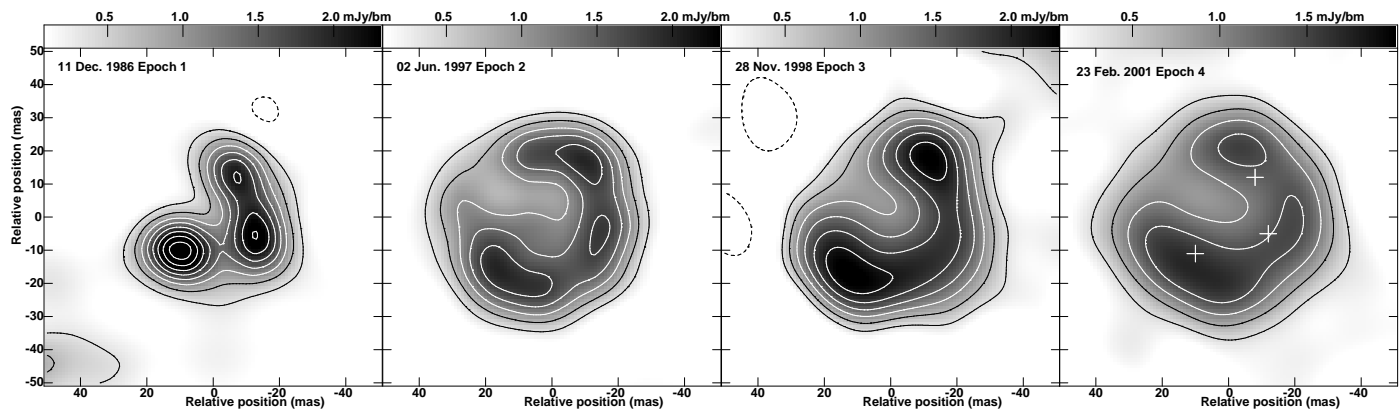}}
\end{picture}
\caption{Contour images from all four VLBI epochs of the compact RSN
43.31$+$592. The date of observation of each epoch is shown at the top
left-hand corner of each image. All four epochs have been convolved
with a circular beamsize of 15\,mas, to match the earlier EVN-only
epochs in 1986 \& 1997. All four images are contoured with -1, 1, 2, 3, 4, 5,
6, 7, 8, 9 and 10 times 0.35\,mJy\,beam$^{-1}$. The grey-scale shown
for three leftmost images is linear and ranges between 0.1 and
2.2\,mJy\,beam$^{-1}$. The grey-scale shown on the 2001 epoch is also
linear but in the range of  0.1 and 1.7\,mJy\,beam$^{-1}$. On the
right-most images three crosses are plotted to show the relative
Gaussian-fitted positions of the three components in epoch 1.}
\label{fig3}
\end{center}
\end{figure*}

\section{Observations}

The latest global VLBI observations were made on the 23 February
2001 using an array with 19 elements at a wavelength of 18\,cm for a
total of 18\,hr. The array used consisted of the ten antennas of
the Very Long Baseline Array (VLBA) in the USA, 7 antennas from
the EVN, a single Very Large Array (VLA) dish and the NASA Deep
Space Network (DSN) antenna at Robledo in Spain. This network was
designed to be as close as possible to that used to make similar
observations on 28 November 1998 by \citet{mcdonald01} which
used a 20 element network. The NASA DSN antenna at Goldstone USA
was not available for the observations presented here, resulting in a
reduction of the number of baselines to 171 compared to the 190
baselines used in the 1998 epoch. Due to the high declination of the
target source, $+$69.7$\degr$, and long duration, 18\,hr, of these observations
the non-availability of the DSN Goldstone antenna did not
significantly compromise the excellent {\it u-v} coverage of this
experiment.

These data were taken in spectral line mode using a total of 128
channels each with a bandwidth of 0.125\,MHz. This
yields a total bandwidth for the observations of 16\,MHz.

Observations of several sources were made in addition to the
main target source which was the central kiloparsec of M82. Throughout
these observations 4.1\,minute scans of M82 where interleaved with 2.1
minute scans of the nearby calibration source
J0958+65 which was used for phase referencing. During the observing
run the bright calibration sources 3C\,84 and J0927+39 were also
observed and used in fringe fitting and bandpass calibration. The source 3C\,84 was also
used in isolation as a flux density scale calibrator.

These observations were correlated using the Mark~IV/VLBA
system at the National Radio Astronomy Observatory (NRAO) in
Socorro, New Mexico. These data were then reduced using the
Astronomical Image Processing System (\textsc{aips}) which is
provided and maintained by the NRAO, following the same
procedures as those detailed in \citet{mcdonald01}.

\begin{figure}
\begin{center}
\setlength{\unitlength}{1mm}
    \begin{picture}(80,65)
\put(0,0){\includegraphics{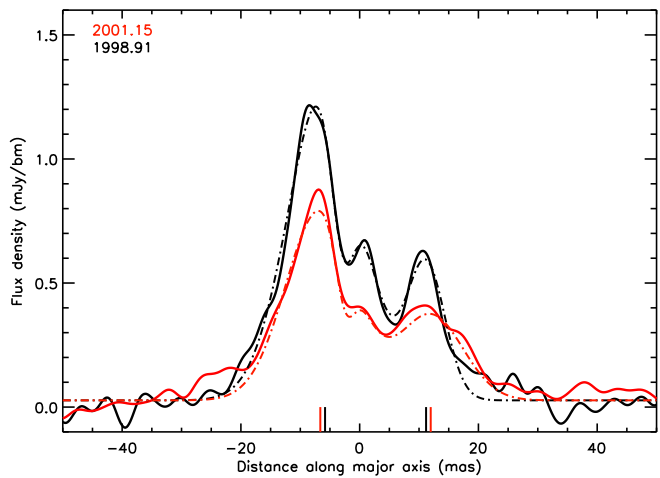}}
\end{picture}
\caption{Two 7.5\,mas wide slices along the major axis of the compact source
41.95+575. The bold black line shows the flux density versus position
for the source in the 1998.91 global VLBI epoch whilst the lighter
grey line is the equivalent slice across the 2001.15 epoch. The
dot-dash line represent the sum of the three-component Gaussian fit to this flux density, with
the positions of the centroids of the two dominant Gaussian components, in each
epoch, marked as extended ticks on the x-axis.  }
\label{41.9-slice}       
\end{center}
\end{figure}

\section{Results \& analysis}

Although the location of all of the bright compact sources were imaged, as in  our earlier study \cite{mcdonald01}
only two were observable with high signal-to-noise, due to the
reduction in brightness sensitivity at high angular resolution. Consequently the results presented here will focus upon the two most
luminous and compact sources in M\,82, 41.95+575 and 43.31+592, and
compare these results with the previous 1.6 and 1.4\,GHz global VLBI
and EVN observations presented in \citet{mcdonald01} and
\citet{pedlar99}. The observational details of each of these epochs
are summarised in Table\,1. In the rest of this section we will
discuss the structural evolution of the two detected compact radio
sources 41.95+575 and 43.31+592.

In Fig.\,1 we present two images of 41.95+575, from both the 28 Nov. 1998 global 
VLBI observations presented by \citet{mcdonald01} and a matched resolution image 
created from this second epoch of global VLBI observations.  Both of these images 
have been created using a Briggs robustness factor \citep{briggs95} of $-$2 and 
convolved to an angular resolution of 3.3\,mas. The measured rms noise of the new 
2001 epoch is 36\,$\mu$Jy\,beam$^{-1}$. The source 41.95+575 is the most compact 
of the sources in M82. This source was only marginally resolved in the earlier 
EVN measurements of \citet{pedlar99}, however both these new VLBI observations 
and those of \citet{mcdonald01} clearly resolve this source showing it to exhibit 
an elongated, bi-polar, structure unlike the shell-like structure commonly 
associated with SNR. The total flux density of 41.95+575 (including the diffuse 
emission surround the two compact components) has evolved between these two 
global VLBI epochs from 34.8$\pm$ 0.1\,mJy in Nov. 1998 to 28.9$\pm$0.3\,mJy in 
Feb. 2001. 

The second most compact radio source within M82 is the shell-like remnant 
43.31+592.  Fig.\,\ref{fig2} shows images from both of the global VLBI 
observations of 43.31+592. Unlike 41.95+575, this source has a well defined shell 
structure and is much more typical of an SNe remnant. Although many of the 
features of the shell are common to both images, at the resolution of these 
global VLBI observations (few milliarcsec) the combination of structural 
evolution of the ring and image fidelity limitations resulting from the 
incomplete sampling of the {\it u-v} plane and non-linear effects of 
deconvolution, result in ambiguities when tracking individual features between 
the two epochs. Hence these images are also presented convolved with a 15\,mas 
circular beam (Fig.\,\ref{fig3}). At this lower angular resolution the effect of 
individual low signal-to-noise components is minimised; this also allows the two 
global VLBI observations to be directly compared with the early EVN-only 
observations made in 1986 and 1997 \citep{pedlar99}.


\begin{table*}
\centering 
\caption{Fit of two Gaussians to the source 
41.95+575 for each of the global VLBI epochs (epoch 3, 1998.91 and
epoch 4, 2001.15). We have not
included measurements derived from the earlier EVN-only observations
in which this source is not resolved.  The quoted flux densities, both peak and
total are derived from the Gaussian fits of the two components and do
not include the diffuse radio emission. The Right Ascensions and
Declinations quoted are relative to 09$^{\rm h}$55$^{\rm m}$,
69\degr40\arcmin (J2000). It should be noted that the absolute
astrometric precision of these positions has been degraded due to the
use of self-calibration, although the relative precision at each
epoch has been preserved.}
\label{tab41.9} 
\begin{tabular}{ccccccccccc}
\hline\noalign{\smallskip}
Epoch &Gaussian & Size & PA & RA & Dec &S$_{\rm peak}$& S$_{\rm tot}$ & \multicolumn{2}{c}{Distance Between Peaks} \\
      &comp.& (mas) & (deg) & (J2000) & (J2000) & (mJy/bm)& (mJy) & (mas) & (pc)\\
\hline\noalign{\smallskip}
&1& 12.7$\times$7.6 &23$\pm$2 &50.68813$\pm$0.00002 &43.7764$\pm$0.0001 &1.80$\pm$0.05&17.9$\pm$0.5  & & \\
\raisebox{1.5ex}[0pt]{3}      &2& 16.2$\times$6.6 &21$\pm$1&50.68529$\pm$0.00002 &43.7682$\pm$0.0002
&1.38$\pm$0.04&15.2$\pm$0.5  &\raisebox{1.5ex}[0pt]{16.6$\pm$0.2}
&\raisebox{1.5ex}[0pt]{0.22$\pm$0.03} \\

\hline\noalign{\smallskip}
                              &1&15.7$\times$11.3  &47$\pm$3  &50.68789$\pm$0.00002  &43.7672$\pm$0.0001  &1.14$\pm$0.02&12.9$\pm$0.4  &&\\
\raisebox{1.5ex}[0pt]{4}      &2&20.2$\times$11.2  &36$\pm$2  &50.68502$\pm$0.00004  &43.7583$\pm$0.0002  &0.71$\pm$0.02&9.1$\pm$0.5  &  \raisebox{1.5ex}[0pt]{17.7$\pm$0.2}  &\raisebox{1.5ex}[0pt]{0.24$\pm$0.03} \\

\noalign{\smallskip}\hline
\end{tabular}
\end{table*}

\begin{table*}
\centering
\caption{Derived radii of SNR shell 43.31$+$592. The first two
columns list radii of peak and 50\% level of the flux density when
averaged within concentric annuli. The flux density versus radius
distribution for each epoch is graphically shown in Fig\,5. The final
column shows radii measured for each epoch derived using the
analytical `Common Point Method' which is described in more detail in
Section 3.3.1. In each of these cases the values listed have been
derived from the matched, 15\,mas angular resolution images of 43.31+592.}
\label{tab2}       
\begin{tabular}{ccccc cc}
\hline\noalign{\smallskip}
&\multicolumn{2}{c}{Annuli (at peak)}&\multicolumn{2}{c}{Annuli (at 50\% of peak)}&\multicolumn{2}{c}{CPM}\\
Epoch&Derived radius&Derived radius&Derived radius&Derived radius&Derived radius&Derived radius\\
 & (mas) & (pc) & (mas) & (pc) & (mas) & (pc)\\

\noalign{\smallskip}\hline\noalign{\smallskip}
1 & $13.4 \pm 1.0$ & $0.179\pm0.013$ & $22.2\pm1.0$ & $0.296\pm0.013$ & $21.0 \pm 0.4$ & $0.326 \pm 0.006$\\
2 & $19.9 \pm 1.0$ & $0.265\pm0.013$ & $29.8\pm1.0$ & $0.397\pm0.013$ & $29.4 \pm 0.6$ & $0.456 \pm 0.009$\\
3 & $20.1 \pm 1.0$ & $0.268\pm0.013$ & $30.5\pm1.0$ & $0.407\pm0.013$ & $31.9 \pm 0.6$ & $0.495 \pm 0.009$\\
4 & $20.4 \pm 1.0$ & $0.272\pm0.013$ & $32.4\pm1.0$ & $0.432\pm0.013$ & $33.1 \pm 0.6$ & $0.514 \pm 0.009$\\
\noalign{\smallskip}\hline
\end{tabular}
\end{table*}

\subsection{The expansion of 41.95$+$575}

From our earlier measurements it was clear that the expansion 
velocity of this object was relatively low, and hence 
measuring the expansion velocity 
of this object over a two year baseline is particularly challenging.
Furthermore the  problem is exacerbated by  changes
in the small scale structure which give rise to problems in
finding common reference points between the two epochs.

\begin{figure}
\begin{center}
\setlength{\unitlength}{1mm}
    \begin{picture}(80,65)
\put(0,0){\includegraphics{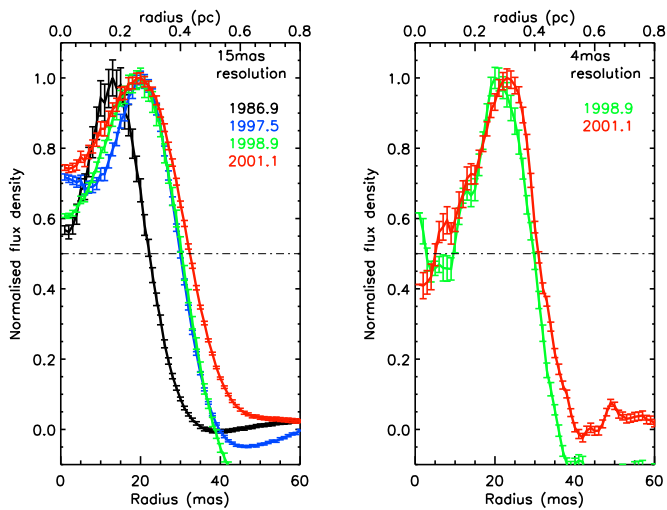}}
\end{picture}
\caption{ Radial flux density profile of the SNR 43.31+592 for each
epoch of observation. The left-hand panel shows four profiles, for
each epoch, derived from images convolved with a 15\,mas synthesised beam,
whilst the right-hand panel shows similar profiles for the two global
VLBI images, 1998.95 and 2001.15, when convolved with a 4\,mas circular beam.}
\label{iring-vel}       
\end{center}
\end{figure}

Several methods were used in an attempt to detect expansion in 41.95+575, 
including taking slices along the major axis of the source as well as fitting the 
positions of the two peaks with 2-D Gaussian components.  
Figure\,\ref{41.9-slice} shows a flux density distribution of a 7.5\,mas wide, 
one-dimensional, slice along the major axis of 41.95+575. In this figure the 
lower, lighter grey line represents the latest global VLBI epoch (2001.15) and 
the upper solid dark black line the first global VLBI epoch (1998.91). Although 
the flux density evolution of this source between the two epochs is clearly shown 
in this figure, the small expansion of the two peaks is not easily discerned. 
However, a small increase in the separation is measured in the Gaussian fitted 
centroids of the two dominant components (the positions of these centroids are 
shown as extended ticks on the x-axis).  This increased separation is measured to 
be 0.7$\pm$0.2\,mas\,year$^{-1}$, which is equivalent to a projected radial 
expansion rate of 2050$\pm$500\,km\,s$^{-1}$.

In addition to fitting the one-dimensional flux density distribution versus 
distance along the major axis of 41.95+575, we have performed extensive 
two-dimensional Gaussian fits to the two brightest components within the image 
plane. The results of this fitting are shown in Table\,\ref{tab41.9}.  The 
expansion derived from these fitted positions is 1.1$\pm$0.3\,mas over the 2.24 
year baseline implying a radial expansion velocity of 1500$\pm$400\,km\,s$^{-1}$.

The expansion along the major axis has been measured in the past by 
\cite{trotman96}, \cite{mcdonald01} and more recently in \cite{muxlow05b}. The 
results presented here are consistent with these results. It is necessary to note 
that as this source appears to be bipolar in nature the expansion along the major 
axis may be greater than the value quoted here due to orientation effects. 
Consequently, and considering the large errors in measuring the expansion of 
41.95+575 over only a 2.24\,year baseline, we tentatively conclude that this 
source is radially expanding at a rate of $\sim$1500--2000\,km\,s$^{-1}$.

\subsection{The expansion of 43.31$+$592}

A comparison between the 1986 and 1997 EVN images of 43.31+592 \citep{pedlar99} 
showed an obvious increase in size consistent with an expansion velocity of 
$\sim$10000 km\,s$^{-1}$. The continuing expansion of this source is evident in 
Fig.\,3.

More detailed expansion measurements of this source using the full resolution 
Global VLBI data poses problems due to evolution of the small-scale structure 
which did not allow accurate Gaussian fitting to the four knots of emission 
identified in \cite{mcdonald01}. Instead, the centre of the source in the maps 
was found for each epoch and the flux was integrated in a series of $1\, 
\mathrm{mas}$-thick annuli at increasing radii from the centre. The geometric 
centre of 43.31+592 was established in each image as the position which was 
equidistant from Gaussian fitted components within the ring. A plot of the 
integrated flux in each annulus against the distance from the centre is plotted 
in Fig.\,\ref{iring-vel} for all epochs at a matched angular resolution of 
15\,mas and for the two latest epochs of global VLBI observations at a resolution 
of 4\,mas. The radius of the flux density maximum and 50\% of this value for each 
of the 4 epochs, measured at a matched angular resolution of 15\,mas, are listed 
in Table\,\ref{tab2}.  The robustness of these derived radii was tested with 
respect to the chosen centre of the remnant by repeating this exercise for 
several centre positions up to a few beam-widths from the calculated centre. It 
was found that, although the radius of the peak of the radial flux density 
profile significantly changed and smeared when the nominated centre was more than 
a beam-width from the true image centre, the radius as measured at a point 
between $\sim$30--70\% of the peak flux did not vary significantly provided the 
central position deviate by more than a beam-width from the true remnant centre.

As can be seen in both Table\,\ref{tab2} and the left-hand panel of 
Fig.\,\ref{iring-vel}, a distinct expansion is observed in the peak radial flux 
density distribution between the 1986.95 epoch and latter epochs in the lower 
(15\,mas) resolution data. However little discernible expansion of the peak flux 
density is recorded over the shorter time lines between the three latter epochs, 
which is not unexpected since an assumed expansion velocity of 
10000\,km\,s$^{-1}$ would imply an increase in the peak radius of $\sim$2.5\,mas 
over the 3.7\,yrs between 1997.42 and 2001.15. Using the higher 4\,mas angular 
resolution global VLBI data, the measured peak of the radial flux density 
distribution (Fig\,\ref{iring-vel}) in 1998.91 is 19.5\,mas compared to 22.5\,mas 
in 2001.15. This implies an angular expansion speed of 
1.34$\pm$0.32\,mas\,yr$^{-1}$, corresponding to a radial expansion speed of 
7800$\pm$1900\,km\,s$^{-1}$.

\begin{figure}
\begin{center}
\includegraphics[width=8.5cm]{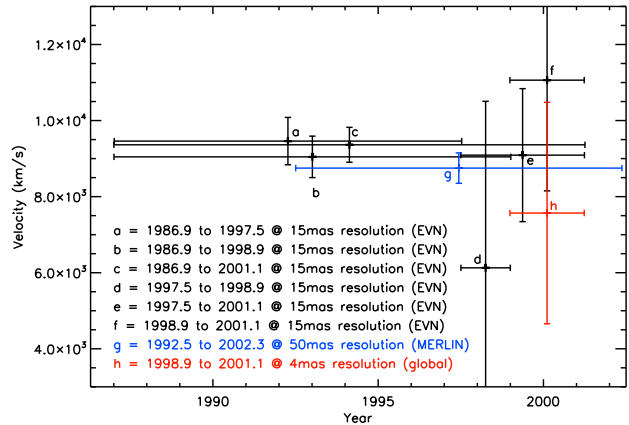}
\caption{Derived velocities for the SNR 43.31+592. Horizontal
bar represents the time baseline between size measurement, vertical bar is the error in measuring the shell radius at 50\% peak flux density
value. Velocities measured between all available epochs have been
included. An additional independent measure (g) of the velocity of
43.31+592 by Muxlow {\it et al.,} (2005), using MERLIN, is also
included for completeness.}
\label{iring}       
\end{center}
\end{figure}

An alternative measure of the radial flux density distribution is the radius at a 
percentage of the peak flux. In Table\,\ref{tab2}, in addition to the peak flux 
density radius, we quote the half peak flux density radius for each epoch of 
observations as measured from the lower 15\,mas resolution images. Whilst neither 
of these values is a true representation of the radius of the SNR's radio 
emission, both are directly proportional to this value and can be used to derive 
the expansion velocity of the SNR (this is discussed in further detail in the 
following section).  In Fig.\,\ref{iring} radial expansion velocities between 
each epoch of observations have been plotted, along with an independent 
measurement of the velocity of 43.31+592 made by comparing deep Multi-Element 
Radio Linked Interferometer Network (MERLIN) 5\,GHz observations made in 1992 and 
2002 (\citealt{muxlow05}). The mean radial expansion velocity of these VLBI 
measurements is 9025$\pm$380\,km\,s$^{-1}$ compared to 
8750$\pm$400\,km\,s$^{-1}$, recorded over a similar period using MERLIN.  These 
results are consistent with the results in \cite{mcdonald01}, \cite{pedlar99} and 
\cite{muxlow05}.

\subsection{The expansion of 43.31+592 measured using an analytical method}

As well as using the conventional methods to measure shell sizes, we have also 
applied a novel analytical technique developed to parameterise the expansion of 
SN1993J in M81 \citet{marcaide06}.

When trying to analyse the expansion of a shell-like object it is important to 
take into account the continuously increasing ratio between the size of such a 
shell and the (constant) beam used for the convolution of the CLEAN components to 
obtain the final image. As the source is expanding, the ratio of the convolved 
beam size and the source size will become proportionally smaller, this fact can 
produce biases in the determination of the expansion parameters. Whilst methods 
for the correction of this beam-related bias in the imaging analysis have been 
developed by \cite{marcaide97}, such solutions tend to produce other undesirable 
new biases in the data. However, a new method for the determination of the 
expansion of radio-emitting shell-like objects that solves this beam bias has 
been developed by \cite{marcaide06}.  When applied to shell-like objects, the 
method has been shown to work in a way totally independent of the a priori beam 
used. This method, called the `Common Point Method' (CPM), is based on 
mathematical properties of the convolution of angular symmetric images with 
Gaussian components. This method is described in detail in \cite{marcaide06}.

\begin{figure}
\includegraphics[width=8cm]{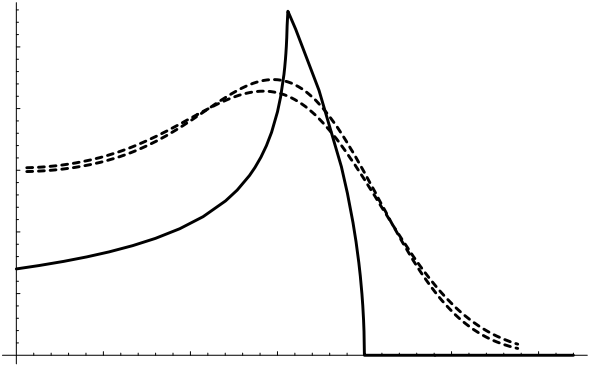}
\caption{Qualitative representation of the properties on which the CPM 
is based. The solid line represents the radial profile of a given
shell-like emitter, whilst the dashed lines show profiles of the angular average of the convolution of 
this shell with two different beams, one of which is 50\% wider than the 
other. The outer point where the dashed curves coincide is very stable 
under beam changes and, thus, can be found to be related with the 
radius of the source within a given iterative process. Note also that
the maxima of both dashed curves is at a radius smaller than the true source radius.}
\label{fig6}       
\end{figure}

In essence, the CPM exploits the fact that there is a point in the angular 
average of such circularly symmetric images that is stable with respect to 
changes in the convolving beam (see Fig.\,\ref{fig6}). This stability can be used 
in an iterative process in order to obtain a magnitude that will be perfectly 
proportional to the radius of the source, with a proportionality constant close 
to unity and will remain constant provided the expansion is nearly self-similar.

The results from applying the CPM to these 43.31+592 data are listed in 
Table\,\ref{tab2} and are in good agreement with the other expansion measurements 
presented in this paper. The measured radial expansion velocity of the remnant 
43.31+592 using the CPM is $11000 \pm 1000$\,km\,s$^{-1}$, fitting to each of the 
four epochs and assuming a free expansion model ({\it i.e.} linear expansion). 
Hence if the expansion is linear, the supernova explosion took place in 
$\sim$1962.

It is possible to fit a wide range of deceleration parameters ($\delta$) to these 
four derived shell radii, since the birth date of this source is unknown. 
Following \cite{mcdonald01} and \cite{huang94} we parameterise the size evolution 
of the remnant as, $D=kT^\delta$, where $D$ and $T$ are the size and age of the 
remnant respectively, $k$ is a constant and $\delta$ is the deceleration 
parameter. In Fig.\,\ref{fig5} the derived shell radii are plotted with various 
fitted parameterised deceleration curves overlaid.  It should be noted that in 
1972 (marked by the vertical dashed line in Fig.\,\ref{fig5}), 43.31+592 was 
detected in radio images made by \cite{kronberg75}. This early radio detection of 
this source implies a lower limit on $\delta$ of 0.68.

\section{Discussion}

From these global VLBI measurements a radial expansion velocity of 
1500$\pm$400\,km\,s$^{-1}$ is derived for the most compact radio source in M82, 
41.95+576. This source is, however, rather anomalous on account of both its high 
radio luminosity (compared to the other supernova remnants within M82), rapid 
flux density decay and highly elongated, bipolar, radio structure. In particular 
the radio structure of 41.95+576 is highly atypical for a SNR. As yet it is 
uncertain what this source is, however it is clear that it is inappropriate to 
treat it as a typical SNR. The nature of 41.95+576 is discussed in 
\cite{muxlow05b}.

\begin{figure}
\includegraphics[width=8.5cm]{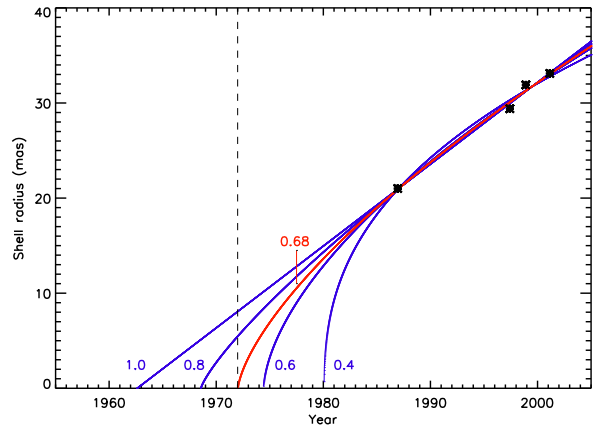}	
\caption{Measured radius of 43.31+592, using the CPM, for the
four epochs. We also show five solid lines representing fitted models
with fixed expansion parameters ranging from 0.4 to 1.0 (free
expansion). The vertical dashed line marks the epoch of the first
detection of this source in 1972 by Kronberg \& Wilkinson (1975).}
\label{fig5}       
\end{figure}

The well-defined shell structure of 43.31+592 is more typical of SNR and radio 
supernovae. As can be seen in Fig.\,\ref{fig5} the present observations are 
consistent with near-free expansion of the shell remnant 43.31+592 with a 
velocity close to 10000\,km\,s$^{-1}$.  This expansion velocity is more than an 
order of magnitude larger than that predicted by \citet{chevalier01} which was 
largely a consequence of the high density and pressure assumed for the M82 ISM. 
The minimum deceleration parameter that is consistent with this expansion and the 
first known recording of the radio source in 1972 \citep{kronberg75} is 0.68, 
which is close to the $\gtsim$0.73 limit placed on this parameter from earlier 
VLBI observations by \cite{mcdonald01}.

\subsection{Interstellar medium pressures in M82}
\label{sec:6}

\cite{chevalier01} assume that the pressure in
the starburst region of M82 is $10^7$\,cm$^{-3}$\,K and from this
predict low SNR expansion velocities of $\sim$500\,km\,s$^{-1}$. While
there is no doubt that part of the starburst must be at this
pressure ({\it e.\,g.} compact H{\sc ii} regions \cite{mcdonald01}), it seems
likely that much of the starburst is not in static pressure
equilibrium and that a range of pressures are present.

As the photoionised component in M82 has a temperature of $\sim$ 10$^4$\,K, the 
pressure assumed by Chevalier \& Fransson would require that the density of this 
gas be $\sim$10$^3$\,cm$^{-3}$. However, observations of free-free absorption 
against individual remnants in M82 give emission measures of $\sim$5$\times 
10^5$\,pc\,cm$^{-6}$ \citep{wills97}. Consequently if the gas number densities 
were to be $\sim$10$^3$\,cm$^{-3}$ the observed emission measure would imply 
ionised gas path lengths of only $\sim$0.5\,pc. Given that the extent of the 
starburst region is at least 500\,pc, unless the ionised gas has contrived to 
occupy only a thousandth of the volume of the starburst in the immediate vicinity 
of each remnant, it is difficult to see how the ionised component can provide 
sufficient pressure to slow the SNR expansion to 500\,km\,s$^{-1}$.

Chevalier \& Fransson suggest that the SNR may be confined by the interclump 
medium of molecular clouds in M82 and note that many of the remnants are in the 
line of sight to regions of strong CO emission. This is, however, not conclusive 
evidence that the remnants are embedded within the clouds, and often the H{\sc i} 
absorption spectra measured directly against the remnants 
\citep{wills98phd,wills98} do not support this interpretation (see discussion of 
43.31+592 below). Furthermore, studies by Wei{\ss}, Walter \& Neininger (1999) 
\nocite{weiss99} have shown that the molecular gas in the starforming regions of 
M82 has a kinetic temperature of $\sim$150\,K and number densities of $\sim 
10^3$\,cm$^{-3}$ which corresponds to a pressure two orders of magnitude less 
than that assumed by Chevalier \& Fransson.

Clearly the ISM of M82 is complex, and to assign a single pressure to the 
starburst region may be unrealistic. Hence it is likely that most of the SNR in 
M82 are embedded in regions with pressures significantly lower than 
$10^7$\,cm$^{-3}$\,K, especially since pre-supernovae stellar winds from the 
massive progenitor stars of these remnants are likely to have driven away much of 
the material within the vicinity of these young supernova remnants prior to the 
supernova phase beginning.  Consequently, it is highly plausible that the 
observed remnants are expanding into regions of the ISM with lower density than 
the average density and hence have velocities significantly in excess of 
500\,km\,s$^{-1}$.

\subsection {Is 43.31+592 in low density region?}

The size, radio luminosity and lack of rapid flux variability of the source 
43.31+592 appears to be typical of young SNRs. In addition it shows a 
well-defined shell structure (Fig.\,\ref{fig2}, \ref{fig3}).  As the shell seems 
to be in, or close to, free expansion (see Section\,3.3) a simple constraint of 
$<2000$\,atoms\,cm$^{-3}$ can be set to the external density by assuming that the 
mass of gas swept up is less than the mass ejected. The expansion velocity of 
this source clearly exceeds the 500\,km\,s$^{-1}$ as discussed above.

This possibility is supported by the lack of a low frequency turnover at 73\,cm 
\citep{wills97} in the spectrum of 43.31+592 even though it is in the line of 
sight to regions with strong free-free emission. This is consistent with the 
remnant being in front of, and not embedded in, the relatively dense ionised gas 
in the central region of M82. Additionally, even though relatively strong H{\sc 
i} absorption ($5 \times 10^{21}$\,atoms\,cm$^{-2}$) is seen against this remnant 
\citep{wills98phd}, the absorption has a relatively narrow width and the systemic 
velocity is consistent with absorption by a disk of H{\sc i} external to the 
central region of M82. Hence this could be further evidence that this source is 
not embedded in neutral gas in the central region of the starburst. Thus if this 
SNR is in a particularly low pressure region, it might possibly reconcile the 
difference between the theoretically predicted and measured expansion velocities.

However it seems unlikely that such a scenario can be contrived to account for 
high expansion velocities in more than a few of the M82 SNR. Recent MERLIN 
observations (\citealt{muxlow05}; \citealt{fenech05}) have independently 
confirmed the expansion velocity for 43.31+592 in addition to expansion 
velocities of several more remnants. Of these other remnants their observed 
radial expansion velocities range from $\sim$2000\,km\,s$^{-1}$ to over 
10000\,km\,s$^{-1}$, implying that neither the relatively slow expansion velocity 
of 41.95+575 nor the much higher velocity of 43.31+592 are particularly atypical.

Hence, we see no reason to revise the supernova rates or remnant ages that we 
have inferred previously. However, we can offer no simple explanation for the 
lack of variability seen in the compact sources \citep{kronberg85}. Over the next 
decade any variability will be constrained by new observations using the next 
generation of high sensitivity radio interferometers, such as the EVLA and {\it 
e}-MERLIN.

\section{Conclusions}

Using these global VLBI observations, the two most compact radio components 
within the centre of the M82 starburst have been imaged with angular resolutions 
of a few milliarcseconds. This latest epoch of observations, in conjunction with 
our previous VLBI observations of M82, complete a $\sim$15 year timeline of VLBI 
observations of these compact sources.

The most compact radio source in M82, 41.95+575, has been imaged using both these 
latest data and those previously published in \cite{mcdonald01}. This compact 
radio source has an elongated, bipolar structure that is atypical of a SNR. 
Between these two global VLBI observations separated by 2.2 year, this source has 
grown with a radial expansion velocity along its major axis and in the plane of 
the sky of 1500$\pm$400\,km\,s$^{-1}$. Between these two global VLBI observations 
the total flux density (including both of the two compact components and the 
surrounding diffuse emission) of 41.95+575 has reduced at a rate of 7.1$\pm0.5$\% 
per year, slightly lower than the previously reported 8.5\% per year 
\citep{trotman96}.

The shell-like remnant 43.31+592 has been imaged with synthesised beams of 4 and 
15\,mas, affording comparisons with both the previous global VLBI observations of 
\cite{mcdonald01} and the earlier EVN-only observations of \cite{pedlar99}. 
Various methods have been used to derive the expansion of this radio shell, all 
of which provide consistent radial velocities of $\sim$10000\,km\,s$^{-1}$ 
between each of the four independent observations.  This value is consistent with 
expansion measurements of this source from observations using other radio 
interferometers \citep{muxlow05}.

Over the $\sim$15 year timeline of these observations 43.31+592 appears to be 
evolving at a rate close to that expected for free-expansion. Using the earliest 
definite radio detection of this compact source in 1972 \citep{kronberg75} these 
observations imply a lower limit on the deceleration parameter of 0.68. Continued 
VLBI observations over an increased time baseline will determine the deceleration 
rate of this source, as well as allowing the interaction of the SNR shell with 
its surrounding ISM to be further investigated.

Higher sensitivity observations using {\it e}-MERLIN and broadband VLBI, combined 
with an increasing time baseline, will enable the expansion velocities of many 
more remnants in M82 to be determined and hence constrain both the supernova 
remnant parameters and the properties of the interstellar medium in M82.

\subsection*{Acknowledgments}
 
We thank the referee for constructive comments. RJB acknowledges financial 
support by the European Commission's I3 Programme ``RADIONET'' under contract No. 
505818. JDR, DF \& MKA acknowledge financial support via PPARC funded 
studentships. The VLBA is operated by the National Radio Astronomy Observatory 
which is a facility of the National Science Foundation operated under cooperative 
agreement by Associated Universities, Inc. The European VLBI Network is a joint 
facility of European, Chinese, South African and other radio astronomy institutes 
funded by their national research councils.

\bibliographystyle{mnras}

\end{document}